\documentstyle[epsf,twoside,fleqn,espcrc2]{article}
\newskip\defaultbaselineskip\defaultbaselineskip=12pt

\def\GeV{\mathord{\rm \;GeV}}

\def\bar{\overline}

\def\eqinf{\mathrel{\raise2pt\hbox to 24pt{\raise -8pt\hbox{$t \to \infty$}\hss{$=$}}}}

\def\gsim{\mathrel{\raise2pt\hbox to 8pt{\raise -5pt\hbox{$\sim$}\hss{$>$}}}}
\def\rsim{\mathrel{\raise2pt\hbox to 8pt{\raise -5pt\hbox{$\sim$}\hss{$>$}}}}
\def\lsim{\mathrel{\raise2pt\hbox to 8pt{\raise -5pt\hbox{$\sim$}\hss{$<$}}}}
\def\ssqr#1#2{{\vbox{\hrule height.#2pt
      \hbox{\vrule width.#2pt height#1pt \kern#1pt\vrule width.#2pt}
      \hrule height.#2pt}\kern-.#2pt}}

\def\Dsl{\,\raise.15ex\hbox{$/$}\mkern-13.5mu D} 
\def\dsl{\raise.15ex\hbox{$/$}\kern-.57em\hbox{$\partial$}}

\ifx\href\undefined\def\href#1#2{{#2}}\fi
\def\spireshome{http://www.slac.stanford.edu/cgi-bin/spiface/find/hep/www?FORMAT=WWW&}
{\catcode`\%=12
\xdef\spiresjournal#1#2#3{\noexpand\protect\noexpand\href{\spireshome
                          rawcmd=find+journal+#1%2C+#2%2C+#3}}
\xdef\spireseprint#1#2{\noexpand\protect\noexpand\href{\spireshome rawcmd=find+eprint+#1%2F#2}}
\xdef\spiresreport#1{\noexpand\protect\noexpand\href{\spireshome rawcmd=find+rept+#1}}
\xdef\spireskey#1{\noexpand\protect\noexpand\href{\spireshome key=#1}}
}

\def\nohref{}

\def\putpaper{\edef\refpage{\the\count0}%
              \def\nohref{}%
              {\def\ {+}\def\nohref##1{}\edef\temp{\noexpand\spiresjournal
               {\journalname}{\volume}{\refpage}}\expandafter}\temp
               {\sfcode`\.=1000{\journalname} {\bf \volume} (\refyear)
                \refpage}\egroup}
\def\putpage{\edef\refpage{\the\count0}%
              \def\nohref{}%
              {\def\ {+}\def\nohref##1{}\edef\temp{\noexpand\spiresjournal
               {\journalname}{\volume}{\refpage}}\expandafter}\temp
              {\refpage}\egroup}
\def\dojournal#1#2 (#3){\def\journalname{#1}\def\volume{#2}\def\refyear
                        {#3}\afterassignment\putpaper\bgroup\count0=}
\def\morepage{\afterassignment\putpage\bgroup\count0=}

\def\NPB#1{\dojournal{Nucl.\ Phys.}{B#1}}

\def\PRL#1{\dojournal{Phys.\ Rev.\ Lett.}{#1}}

\def\PRD#1{\dojournal{Phys.\ Rev.}{D#1}}

\def\etal{{\it et al.}}

\newcommand{\AmS}{{\protect\the\textfont2
  A\kern-.1667em\lower.5ex\hbox{M}\kern-.125emS}}

\def\GeV{\mathord{\rm \;GeV}}

\def\NDR{\mathord{\rm \;NDR}}
\def\vev#1{\langle      #1 \rangle}

\hyphenation{author another created financial paper re-commend-ed}

\title{Wilson versus Clover fermions: A case for improvement}

\author{Rajan Gupta\address{Group T-8, MS B-285, Los Alamos National Laboratory, %
        Los Alamos, New Mexico, 87545, USA}%
        \thanks{In collaboration with T. Bhattacharya and S. Sharpe.}
}
       
\begin{document}

\begin{abstract}
We present evidence for improvement with tadpole improved clover
fermions based on an analysis of the chiral behavior of $B_K$ and the
quark condensate.  Also presented are a comparison of the mass
splittings in the baryon octet and decuplet, a calculation of $c_A$
using standard 2-point correlation functions, and the problem of zero
modes of the Dirac operator.
\end{abstract}

\maketitle

\section{$B_K$}

With Wilson fermions, straightforward calculations of $B_K$ using the
1-loop improved $\Delta S = 2$ operator fail due to the large mixing
with the wrong chirality operators~\cite{bkw97lanl}. Since this mixing
is an artifact of lattice discretization, one hopes that it can be
significantly reduced by improving the action. By comparing results
obtained using the Wilson and the tadpole improved clover action
($c_{SW}=1.4785$) on the same quenched gauge lattices (170 lattices of
size $32^3 \times 64$ at $\beta = 6.0$) we show that this is indeed
the case.

\begin{figure}[t]
\vspace{9pt}
\hbox{\hskip15bp\epsfxsize=0.9\hsize \epsfbox {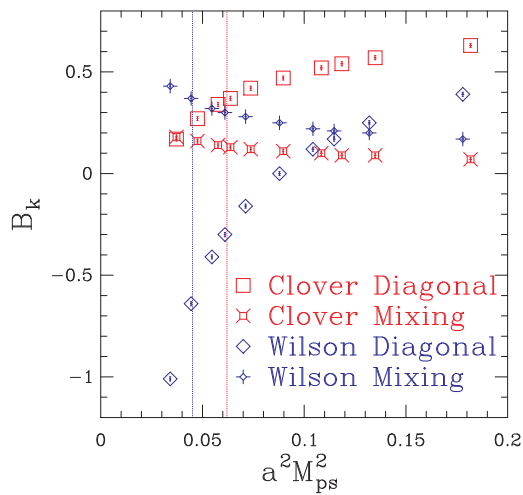}}
\vskip -0.8cm
\caption{Data for $B_K$ for Wilson and clover fermions. The diagonal
($\Delta S = 2$) and mixing operators contribution are shown
separately.  Vertical lines are at $M_K =495$ MeV.}
\vskip -0.6cm
\label{f:Bkw}
\end{figure}

Fig.~\ref{f:Bkw} shows the Wilson and clover data as
a function of $a^2M_K^2$. For each data set, $B_K$ is written as the
sum of two parts $--$ the contribution of the diagonal (the 1-loop
tadpole improved $LL$) operator, and the mixing term which is
proportional to $\alpha_s$.  The general form, ignoring chiral
logarithms and terms proportional to $(m_s-m_d)^2$, for $p_i=p_f=M_K$
is~\cite{bkw97lanl}
\begin{eqnarray*}
& & B_K(M_K) = {\left\langle \overline{K^0}(p_f)
\right| {\cal O} \left| K^0(p_i) \right\rangle
\over   (8/3) f_{K}^2 M_K^2} \\ 
 &=& {\alpha \over M_K^2} + \beta + \gamma + 
   (\delta_1 + \delta_2 + \delta_3) M_K^2 + \ldots .
\label{eq:Bkcpt}
\end{eqnarray*}
The coefficients $\alpha, \beta, \delta_1$ are pure artifacts,
therefore their value can be used to quantify improvement.  Of these
$\alpha$ is the most serious as it causes $B_K$ to diverge in the
chiral limit.

The divergence, in the limit $M_K \to 0$, of the diagonal term due to
a non-zero $\alpha$ is evident in Fig.~\ref{f:Bkw} for Wilson
fermions.  This artifact is only partially cancelled by the 1-loop
mixing operator. The situation is considerably improved with clover
fermions.  The corresponding values at $M_K = 495$ MeV are $B_K({\rm
Wilson}) = - 0.29(1)$ whereas $B_K({\rm clover}) = 0.50(1)$.  This
improvement arises because the two dominant artifacts $\alpha $ and
$\beta $ are significantly reduced; $\alpha(W) = 0.0460(14)$ versus
$\alpha(C)=-0.0125(6)$, and $\beta(W) = 0.048(46)$ versus
$\beta(C)=-0.006(37)$.

As explained in~\cite{bkw97lanl}, the contributions proportional to
$\alpha, \beta, \delta_1$ can be removed completely by studying the
momentum dependence of the matrix elements. Short of calculating the
mixing coefficients non-perturbatively, the way to remove the
artifacts in $\gamma, \delta_2, \delta_3$ is to extrapolate to
$a=0$. We have done the calculation at $\beta=6.0$ only, where our
final results are $B_K(\NDR,2\GeV) = 0.68(7)$ and $0.58(7)$ for Wilson
and clover formulations respectively.  The benchmark value, including
$a \to 0$ extrapolation, is $B_K({\rm Staggered}) = 0.628(41)$, as
obtained by the JLQCD collaboration~\cite{BK97JLQCD}.

\section{QUARK CONDENSATE}

\begin{figure}[t]
\vspace{9pt}
\hbox{\hskip15bp\epsfxsize=0.9\hsize \epsfbox {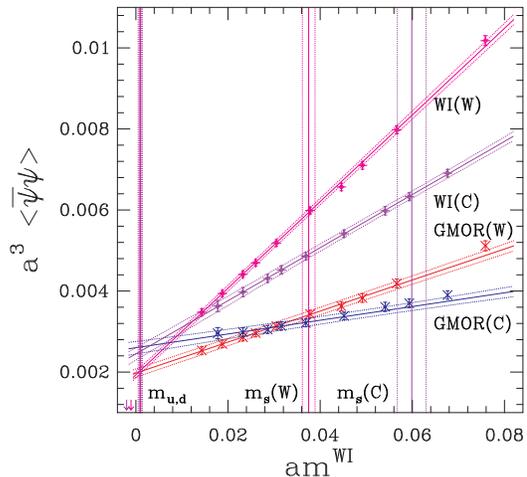}}
\vskip -0.8cm
\caption{Data for $\vev{\bar\psi\psi}$ versus the quark mass for
Wilson (W) and clover (C) fermions.  The vertical lines show $m_{u,d}$
and $m_s$ for the two actions. The $1/a$ are $2.33(4)$ and $1.99(5)$ GeV for 
Wilson and Clover fermions.}
\vskip -0.6cm
\label{f:xbarx}
\end{figure}

The chiral condensate $\vev{\bar \psi \psi}$ is not simply related to
the trace of the Wilson quark propagator $\vev{S_F(0,0)}$. The
breaking of chiral symmetry by the $r$ term introduces contact terms
that need to be subtracted non-perturbatively from
$\vev{S_F(0,0)}$~\cite{WI85Bochicchio}. This has not proven
practical. Instead, the methods of choice are to either evaluate the
right hand side of the continuum Ward Identity 
\begin{eqnarray}
\label{eq:WIcondensate}
\vev{\bar\psi\psi}^{\rm WI} &\equiv& \vev{0|S_F(0,0)|0} \nonumber \\
        &=& \lim_{m_q\to 0} m_q \int d^4x \vev{0|P(x)P(0)|0},
\end{eqnarray}
or cast the Gell-Mann, Oakes, Renner relation
\begin{equation}
\label{eq:GMOR}
\vev{\bar\psi\psi}^{\rm GMOR}
        = \lim_{m_q\to 0} -{ f_\pi^2 M_\pi^2 \over {4 m_q}} \,.
\end{equation}
in terms of lattice correlation functions~\cite{hm92LANL}. These estimates have errors
of both $O(a)$ and $O(ma)$, and at fixed $a$ are therefore expected to
agree only in the chiral limit.  A comparison of the efficacy of the
two methods is shown in Fig.~\ref{f:xbarx}.

We find that a reliable extrapolation to the chiral limit can be made using a
linear fit, and the two methods give consistent results for both
Wilson and clover fermions. Also, the $O(ma)$ corrections are
significantly smaller for clover fermion.

\section{BARYON MASS-SPLITTINGS}

\begin{figure}[t]
\vspace{9pt}
\hbox{\hskip15bp\epsfxsize=0.9\hsize \epsfbox {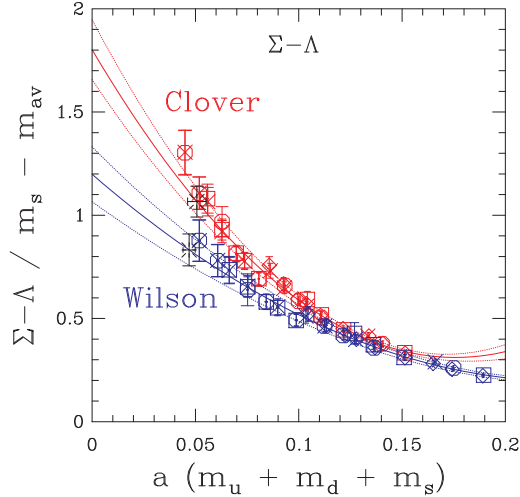}}
\vskip -0.8cm
\caption{Behavior of the $\Sigma - \Lambda$ splitting versus the sum of the three quark masses. The 
leading mass dependence $m_s - (m_u+m_d)/2$ has been divided out.}
\vskip -0.6cm
\label{f:SigLam}
\end{figure}

In Ref.~\cite{HM97LANL} we presented a detailed analysis of
mass-splittings in the baryon octet and decuplet with Wilson
fermions. We had found a large non-linear dependence on quark mass for
the $\Sigma - \Lambda$, $\Sigma - N$, and $\Xi - N$
splittings. Extrapolation of the data to the physical masses including
these non-linearities gave estimates consistent with observed
values. On the other hand we had found a surprisingly good linear fit
to the decuplet masses, and the splittings were underestimated by
$25-30\%$.  The data with clover fermions show the same qualitative
features.  As an illustration, we show a comparison of the $\Sigma -
\Lambda$ splitting in Fig.~\ref{f:SigLam}.  Details of the
analysis will be published elsewhere~\cite{HM99LANL}.

\section{DETERMINATION OF $c_A$}

\begin{figure}[ht]
\vspace{9pt}
\hbox{\hskip15bp\epsfxsize=0.9\hsize \epsfbox {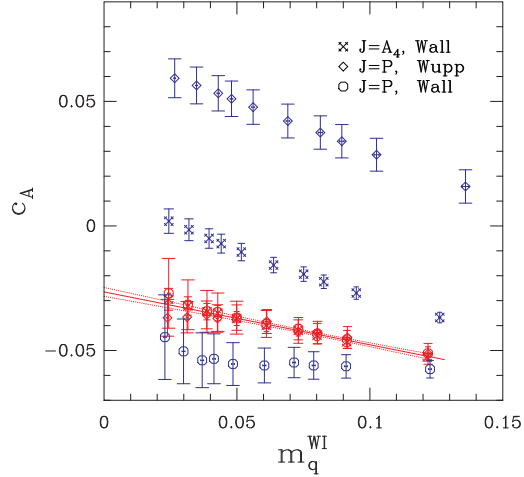}}
\vskip -0.8cm
\caption{Wilson and clover fermion results for $c_A$ as a function of
$(m_1+m_2)$, and for three different sources $J$. The clover data 
is seen to be independent of $J$ and shows a linear behavior.}
\vskip -0.6cm
\label{f:cA}
\end{figure}

The improvement coefficient for the axial current, $c_A$, is
calculated using the the axial WI~\cite{alpha-1}.  If the clover
coefficient $c_{SW} $ is tuned to its non-perturbative value $1.769$
at $\beta=6.0$~\cite{alpha-1}, the sum $(m_1+m_2)$ of quark masses
defined by
\begin{eqnarray*}
  \frac{ \sum_{\vec{x}} \langle
 \partial_\mu [A_\mu + a c_A \partial_\mu P]^{(12)}(\vec{x},t) J^{(21)}(0) \rangle}
 {\sum_{\vec{x}} \langle P^{(12)}(\vec{x},t) J^{(21)}(0) \rangle}
\label{cA} 
\end{eqnarray*}
should be independent of $t$ and the initial pseudoscalar state
created by $J$, up to corrections of $O(a^2)$. We vary the composition
of the initial state by using $J=P$ or $A_4$ and by using ``wall'' or
``Wuppertal'' smearing functions in the calculation of the quark
propagators.  The results in Fig.~\ref{f:cA} show a large dependence
on the initial state for Wilson fermions and almost none already for
$c_{SW}=1.4785$!  We estimate $c_A =- 0.026(2)$ from this clover data,
whereas the ALPHA collaboration report $c_A =-0.083(5)$ at $c_{SW}=
1.769$~\cite{alpha-1}. We are repeating the calculation at $c_{SW}=
1.769$ to understand this difference.

\section{ZERO MODES}

The explicit breaking of chiral symmetry in Wilson-like fermions gives
rise to the problem of ``exceptional configurations'' in the quenched
theory. The cause is that the Wilson $r$ term breaks the
anti-hermitian property of the massless Dirac operator. As a result,
zero modes of the Dirac operator extend into the physical region
$\kappa < \kappa_c$. Thus, on a given configuration, as the quark mass
is lowered and approaches the first of the unphysical modes, one
encounters exceptionally large fluctuations in the correlation
functions. Such configurations dominate the ensemble average and as
discussed in~\cite{Bardeen} there is no basis for excluding
them. Tuning $c_{SW}$ reduces the $O(a)$ chiral symmetry breaking
artifacts as shown above, however, it does not reduce this
problem~\cite{Bardeen}.  We find, by comparing fluctuations in 2-point
and 3-point correlation functions between Wilson and Clover fermions, that
the problem, in fact, gets worse.  A deeper understanding of the
persistence of the zero mode problem even though the chiral behavior
is improved is missing.

\section*{Acknowledgements}

This work was supported by the DoE Grand Challenges award at the
Advanced Computing Lab at Los Alamos, and by the NATO Collaborative
Research Grant, contract no. 940451.

\end{document}